\documentclass[12pt]{article}

\newcommand{\agt}{\,\rlap{\lower 3.5 pt \hbox{$\mathchar \sim$}} \raise 1pt
 \hbox {$>$}\,}
\newcommand{\alt}{\,\rlap{\lower 3.5 pt \hbox{$\mathchar \sim$}} \raise 1pt
 \hbox {$<$}\,}

\begin{document}

\title{
\vskip-3cm{\baselineskip14pt
\centerline{\normalsize DESY 08-021\hfill ISSN 0418-9833}
\centerline{\normalsize TTP08-10\hfill}
\centerline{\normalsize LPSC 08-020\hfill}
\centerline{\normalsize February 2008\hfill}}
\vskip1.5cm
\bf Ghost contributions to charmonium production in polarized high-energy
collisions}

\author{M. Klasen$^{a,}$\thanks{klasen@lpsc.in2p3.fr},
B. A.\ Kniehl$^{b,}$\thanks{kniehl@desy.de},
L. N. Mihail\u a$^{c,}$\thanks{luminita@particle.uni-karlsruhe.de},
M. Steinhauser$^{c,}$\thanks{matthias.steinhauser@uka.de}\\
\\
{\normalsize\it $^a$ Laboratoire de Physique Subatomique et de Cosmologie,}\\
{\normalsize\it Universit\'e Joseph Fourier/CNRS-IN2P3/INPG,}\\
{\normalsize\it 53 Avenue des Martyrs, 38026 Grenoble, France}\\
\\
{\normalsize\it $^b$ II. Institut f\"ur Theoretische Physik, Universit\"a
t Hamburg,}\\
{\normalsize\it Luruper Chaussee 149, 22761 Hamburg, Germany}\\
\\
{\normalsize\it $^c$ Institut f\"ur Theoretische Teilchenphysik,
Universit\"at Karlsruhe,}\\
{\normalsize\it Engesserstra\ss e 7, 76131 Karlsruhe, Germany}}

\date{}

\maketitle

\begin{abstract}
In a previous paper [Phys.\ Rev.\  D {\bf 68}, 034017 (2003)], we investigated
the inclusive production of prompt $J/\psi$ mesons in polarized hadron-hadron,
photon-hadron, and photon-photon collisions in the factorization formalism of
nonrelativistic quantum chromodynamics providing compact analytic results for
the double longitudinal-spin asymmetry $\mathcal{A}_{LL}$.
For convenience, we adopted a simplified expression for the tensor product of
the gluon polarization four-vector with its charge conjugate, at the expense
of allowing for ghost and anti-ghosts to appear as external particles.
While such ghost contributions cancel in the cross section asymmetry
$\mathcal{A}_{LL}$ and thus were not listed in our previous paper, they do
contribute to the absolute cross sections.
For completeness and the reader's convenience, they are provided in this
addendum.
\medskip

\end{abstract}

The factorization formalism of nonrelativistic QCD (NRQCD)
\cite{Caswell:1985ui} provides a rigorous theoretical framework for the
description of heavy-quarkonium production and decay.
This formalism implies a separation of process-dependent short-distance
coefficients, to be calculated perturbatively as expansions in the
strong-coupling constant $\alpha_s$, from supposedly process-independent
long-distance matrix elements (MEs), to be extracted from experiment, and
takes into account the complete structure of the $Q\overline{Q}$ Fock space,
which is spanned by the states $n={}^{2S+1}L_J^{(C)}$ with definite spin $S$,
orbital angular momentum $L$, total angular momentum $J$, and color
multiplicity $C=1,8$.
By velocity scaling rules, the MEs are predicted to scale with a definite
power of the heavy-quark ($Q$) velocity $v\ll1$, so that a small number of
these non-perturbative parameters should allow for meaningful predictions in
practice.

In Ref.~\cite{Klasen:2003zn}, we applied the NRQCD factorization formalism to
the inclusive production of prompt $J/\psi$ mesons in polarized hadron-hadron,
photon-hadron, and photon-photon collisions and provided compact analytic
results for the double longitudinal-spin asymmetry $\mathcal{A}_{LL}$,
defined in Eq.~(2.1) of Ref.~\cite{Klasen:2003zn}.
Specifically, we considered inclusive $J/\psi$ production in polarized $pp$,
$\gamma d$, and $\gamma\gamma$ collisions, appropriate for the RHIC-Spin
experiments at the BNL Relativistic Heavy Ion Collider (RHIC), the SLAC
fixed-target experiment E161, and the TeV-Energy Superconducting Linear
Accelerator (TESLA) operated in the $e^+e^-$ and $\gamma\gamma$ modes,
respectively.
We took the $J/\psi$ mesons to be unpolarized.

There is a technical subtlety related to the definition of the polarization
four-vector $\varepsilon(p,\xi)$ of an external gluon, with four-momentum $p$
and helicity quantum number $\xi=\pm1$, which is potentially prone to create
confusion.
As for the tensor product of $\varepsilon(p,\xi)$ with its charge conjugate, a
natural choice, which avoids the introduction of unphysical degrees of gluon
polarization, is
\begin{equation}
\varepsilon_\mu(p,\xi)\varepsilon_\nu^*(p,\xi)
=\frac{1}{2}\left(-g_{\mu\nu}+\frac{p_\mu\eta_\nu+p_\nu\eta_\mu}{k\cdot\eta}
+i\xi\epsilon_{\mu\nu\rho\sigma}\frac{p^\rho \eta^\sigma}{p\cdot \eta} \right),
\label{eq:1}
\end{equation}
where $\eta$ is an arbitrary light-like four-vector orthogonal to $p$, with
$\eta^2=0\ne p\cdot\eta$.
An obvious disadvantage of Eq.~(\ref{eq:1}) is that it introduces a host of
terms involving $\eta$ in intermediate results.
In practical calculations such as the one performed in
Ref.~\cite{Klasen:2003zn}, it is therefore advantageous to omit the second
term on the right-hand side of Eq.~(\ref{eq:1}) and to identify $\eta$ with
the four-momentum $p^\prime$ of another external parton \cite{Bojak:1998bd},
so that
\begin{equation}
\varepsilon_\mu(p,\xi)\varepsilon_\nu^*(p,\xi)
=\frac{1}{2}\left(-g_{\mu\nu}
+i\xi\epsilon_{\mu\nu\rho\sigma}\frac{p^\rho p^{\prime\sigma}}{p\cdot p^\prime}
\right),
\label{eq:2}
\end{equation}
at the expense of endowing the gluon with unphysical degrees of polarization,
which must be eliminated by subtracting contributions arising from the
presence of its ghost $h$ and anti-ghost $\overline{h}$ as external particles.
Since such ghost contributions cancel in the cross section differences
appearing in the numerator of $\mathcal{A}_{LL}$, as illustrated below, we did
not list them in Ref.~\cite{Klasen:2003zn}.
However, they are necessary to recover the well-known expressions for the
unpolarized cross sections entering the denominator of $\mathcal{A}_{LL}$, as
we did.
In this sense, they were included in our numerical analysis.

Recently, there has been renewed interest in charmonium production by
polarized hadron-hadron and photon-hadron collisions
\cite{Nayak:2005ty,Meijer:2007eb}.
In Ref.~\cite{Nayak:2005ty}, the $J/\psi$ and $\psi^\prime$ polarizations were
predicted for polarized $pp$ collisions at RHIC-Spin.
In Ref.~\cite{Meijer:2007eb}, the squares of the helicity amplitudes
$\mathcal{M}(a,b,c)$ of the partonic subprocesses
$\gamma(a)+g(b)\to Q\overline{Q}[n]+g(c)$ and
$g(a)+g(b)\to Q\overline{Q}[n]+g(c)$ were listed for
$n={}^1S_0^{(C)},{}^3S_1^{(C)},{}^1P_1^{(C)},{}^3P_J^{(C)}$ with
$J=0,1,2$ and $C=1,8$.
The longitudinally-polarized differential cross sections evaluated from these
helicity amplitudes were found to agree with our results \cite{Klasen:2003zn}
after properly subtracting the ghost contributions mentioned above, which we
had provided to the authors of Ref.~\cite{Meijer:2007eb} via private
communication.
Since these contributions may be useful for applications by other authors as
well, we decided to publish them in this addendum to
Ref.~\cite{Klasen:2003zn}.

In the following, we present the differential cross sections
$\mathrm{d}\sigma/\mathrm{d}t$ of the partonic subprocesses
\begin{equation}
\{\gamma,g\}h\to c\overline{c}[n]h.
\end{equation}
Here and in the following, $s$, $t$, and $u$ denote the usual Mandelstam
variables.
The results for $\{\gamma,g\}\overline{h}\to c\overline{c}[n]\overline{h}$ are
identical by charge-conjugation invariance, while those for
$h\{\gamma,g\}\to c\overline{c}[n]h$ and
$h\overline{h}\to c\overline{c}[n]\{\gamma,g\}$ are related by crossing
symmetry, as indicated below.
As usual, $\mathrm{d}\sigma/\mathrm{d}t$ is evaluated from the absolute square
of the transition matrix element $\mathcal{M}$ through multiplication with
factors for flux, phase space, spin, and color, as
\begin{equation}
\frac{\mathrm{d}\sigma}{\mathrm{d}t}=
\frac{1}{2s}\,\frac{1}{8\pi s}\,\frac{1}{4}\left(\frac{1}{8}\right)^i
|\mathcal{M}|^2,
\label{eq:3}
\end{equation}
where $i=1,2$ is the number of color-octet partons (gluons or ghosts) in the
initial state.

The only non-vanishing ghost contributions read
\begin{eqnarray}
|\mathcal{M}|^2(\gamma h\to c\overline{c}[{}^1S_0^{(8)}]h)&=&
\frac{24e^2g_s^4\langle\mathcal{O}[{}^1S_0^{(8)}]\rangle Q_c^2su}{Mt(s+u)^2},
\nonumber\\
|\mathcal{M}|^2(\gamma h\to c\overline{c}[{}^3P_0^{(8)}]h)&=&
\frac{32e^2g_s^4\langle\mathcal{O}[{}^3P_0^{(8)}]\rangle Q_c^2su}{M^3t(s+u)^4}
[(2t+3u)^2+6s(2t+3u)+9s^2],
\nonumber\\
|\mathcal{M}|^2(\gamma h\to c\overline{c}[{}^3P_1^{(8)}]h)&=&
\frac{32e^2g_s^4\langle\mathcal{O}[{}^3P_1^{(8)}]\rangle Q_c^2}{M^3(s+u)^4}
[u^2(t+u)-su^2+s^2(t-u)+s^3],
\nonumber\\
|\mathcal{M}|^2(\gamma h\to c\overline{c}[{}^3P_2^{(8)}]h)&=&
\frac{32e^2g_s^4\langle\mathcal{O}[{}^3P_2^{(8)}]\rangle Q_c^2}{5M^3t(s+u)^4}
[3tu^2(t+u)+su(8t^2+21tu+12u^2)
\nonumber\\
&&{}+3s^2(t^2+7tu+8u^2)+3s^3(t+4u)],
\nonumber\\
|\mathcal{M}|^2(gh\to c\overline{c}[{}^1S_0^{(1)}]h)&=&
\frac{4g_s^2\langle\mathcal{O}[{}^1S_0^{(1)}]\rangle}
{3e^2Q_c^2\langle\mathcal{O}[{}^1S_0^{(8)}]\rangle}
|\mathcal{M}|^2(\gamma h\to c\overline{c}[{}^1S_0^{(8)}]h),
\nonumber\\
|\mathcal{M}|^2(gh\to c\overline{c}[{}^3P_J^{(1)}]h)&=&
\frac{4g_s^2\langle\mathcal{O}[{}^3P_J^{(1)}]\rangle}
{3e^2Q_c^2\langle\mathcal{O}[{}^3P_J^{(8)}]\rangle}
|\mathcal{M}|^2(\gamma h\to c\overline{c}[{}^3P_J^{(8)}]h),
\nonumber\\
|\mathcal{M}|^2(gh\to c\overline{c}[{}^1S_0^{(8)}]h)&=&
\frac{5g_s^2}{12e^2Q_c^2}
|\mathcal{M}|^2(\gamma h\to c\overline{c}[{}^1S_0^{(8)}]h),
\nonumber\\
|\mathcal{M}|^2(gh\to c\overline{c}[{}^3S_1^{(8)}]h)&=&
\frac{3g_s^6\langle\mathcal{O}[{}^3S_1^{(8)}]\rangle}{4M^5stu(s+u)^2}
[tu^2(t+u)^2(3t-u)+su(-2t^4+2t^3u
\nonumber\\
&&{}+7t^2u^2+4tu^3+u^4)
+s^2(3t^4+t^3u-4t^2u^2-3tu^3+u^4)
\nonumber\\
&&{}+s^3(4t^3+7t^2u-2tu^2-u^3)+s^4(t^2+6tu-u^2)],
\nonumber\\
|\mathcal{M}|^2(gh\to c\overline{c}[{}^1P_1^{(8)}]h)&=&
\frac{g_s^2\langle\mathcal{O}[{}^1P_1^{(8)}]\rangle}
{e^2Q_c^2\langle\mathcal{O}[{}^1S_0^{(8)}]\rangle M^2}
|\mathcal{M}|^2(\gamma h\to c\overline{c}[{}^1S_0^{(8)}]h),
\nonumber\\
|\mathcal{M}|^2(gh\to c\overline{c}[{}^3P_J^{(8)}]h)&=&
\frac{5g_s^2}{12e^2Q_c^2}
|\mathcal{M}|^2(\gamma h\to c\overline{c}[{}^3P_J^{(8)}]h),
\label{eq:4}
\end{eqnarray}
where $e=\sqrt{4\pi\alpha}$, with $\alpha$ being Sommerfeld's fine-structure
constant, and $g_s=\sqrt{4\pi\alpha_s}$ are the electromagnetic and strong
gauge couplings, $Q_c$ and $m_c$ are the fractional electric charge and mass
of the $c$ quark, and $M=2m_c$.
By four-momentum conservation, we have $s+t+u=M^2$.

We now explain how the unpolarized and polarized results of
Refs.~\cite{Meijer:2007eb,Yuan:1999eb} may be recovered from the results of
Ref.~\cite{Klasen:2003zn} in combination with Eqs.~(\ref{eq:3}) and
(\ref{eq:4}), considering $\gamma g\to c\overline{c}[{}^1S_0^{(8)}]g$ as an
example.
The unpolarized and polarized results of Eqs.~(A4) and (A5) in
Ref.~\cite{Yuan:1999eb} are obtained from Eq.~(\ref{eq:3}) by inserting
\begin{eqnarray}
|\mathcal{M}|_\mathrm{unpol}^2(\gamma g\to c\overline{c}[{}^1S_0^{(8)}]g)&=&
\sum_{\xi_a,\xi_b=\pm1}
|\mathcal{M}|_{\xi_a,\xi_b}^2(\gamma g\to c\overline{c}[{}^1S_0^{(8)}]g)
-|\mathcal{M}|^2(\gamma h\to c\overline{c}[{}^1S_0^{(8)}]h)
\nonumber\\
&&{}-|\mathcal{M}|^2(\gamma\overline{h}\to c\overline{c}[{}^1S_0^{(8)}]
\overline{h}),
\nonumber\\
|\mathcal{M}|_{LL}^2(\gamma g\to c\overline{c}[{}^1S_0^{(8)}]g)&=&
\sum_{\xi_a,\xi_b=\pm1}(-1)^{\xi_a\xi_b}
|\mathcal{M}|_{\xi_a,\xi_b}^2(\gamma g\to c\overline{c}[{}^1S_0^{(8)}]g),
\end{eqnarray}
respectively,
where
$|\mathcal{M}|_{\xi_a,\xi_b}^2(\gamma g\to c\overline{c}[{}^1S_0^{(8)}]g)$
may be gleaned from Eq.~(A5) of Ref.~\cite{Klasen:2003zn} and
$|\mathcal{M}|^2(\gamma h\to c\overline{c}[{}^1S_0^{(8)}]h)
=|\mathcal{M}|^2(\gamma\overline{h}\to c\overline{c}[{}^1S_0^{(8)}]
\overline{h})$ is given in Eq.~(\ref{eq:4}) above.
As mentioned above, all ingredients entering
$|\mathcal{M}|_{LL}^2(\gamma g\to c\overline{c}[{}^1S_0^{(8)}]g)$ are
contained in Ref.~\cite{Klasen:2003zn}.
By crossing symmetry, we have
\begin{eqnarray}
|\mathcal{M}|^2(h\gamma\to c\overline{c}[{}^1S_0^{(8)}]h)&=&
\left.|\mathcal{M}|^2(\gamma h\to c\overline{c}[{}^1S_0^{(8)}]h)\right|_{t
\leftrightarrow u},
\nonumber\\
|\mathcal{M}|^2(h\overline{h}\to c\overline{c}[{}^1S_0^{(8)}]\gamma)&=&
\left.|\mathcal{M}|^2(\gamma h\to c\overline{c}[{}^1S_0^{(8)}]h)\right|_{s
\leftrightarrow t}.
\end{eqnarray}
Similar relationships hold for the other partonic subprocesses involving two
external gluons considered in Ref.~\cite{Klasen:2003zn}.

In conclusion, we complemented the partonic cross sections for the inclusive
production of prompt $J/\psi$ mesons in polarized hadron-hadron,
photon-hadron, and photon-photon collisions listed in the Appendix of
Ref.~\cite{Klasen:2003zn} by providing the ghost contributions, which cancel
in the cross section differences entering $\mathcal{A}_{LL}$, but contribute
to absolute cross sections, including the unpolarized ones.

We thank Jack Smith for carefully comparing the results of
Ref.~\cite{Meijer:2007eb} with ours \cite{Klasen:2003zn}.
The work of B.A.K. was supported in part by the German Federal Ministry for
Education and Research BMBF through Grant No.\ 05~HT6GUA, by
the German Research Foundation DFG through Grant No.\ KN~365/6--1, and by
the Helmholtz Association through Grant No.\ HA~101.

\end{document}